\documentclass[12pt]{article}
\usepackage{amsthm,amsmath,latexsym,amssymb,amsfonts,amscd}
\usepackage{graphics,lscape,fancyhdr,array,stmaryrd,euscript,wrapfig}
\usepackage{verbatim,slashed,empheq}
\numberwithin{equation}{section}
\usepackage{hyperref,setspace}

\usepackage{mathrsfs}
\usepackage[numbers,sort&compress]{natbib}
\usepackage[nottoc]{tocbibind}

\newcommand{\pl}{\partial}

\newcommand{\bref}[1]{\textbf{\ref{#1}}}
\newcommand{\be}{\begin{equation}}
\newcommand{\ee}{\end{equation}}

\newcommand{\mm}{{\ensuremath{{\mu}}}}

\newcommand{\aA}{{\ensuremath{\mathcal{A}}}}
\newcommand{\aB}{{\ensuremath{\mathcal{B}}}}

\newcommand{\bry}{{{\bar{y}}}}

\newcommand{\fud}[2]{{}^{#1}{}_{#2}\,}
\newcommand{\fdu}[2]{{}_{#1}{}^{#2}\,}

\newcommand{\hs}{{\mathfrak{hs}}}
\newcommand{\tr}{{\mathrm{tr}}}

\newcommand{\action}[2]{{\left\langle\vphantom{#2}#1\,\right|\left.\vphantom{#1}#2\right\rangle}}
\newcommand{\scalar}[2]{{\left\langle\vphantom{#2}#1\right.;\left.\vphantom{#1}#2\right\rangle}}

\newcommand{\besubeqs}{\begin{subequations}}
\newcommand{\esubeqs}{\end{subequations}}

\usepackage{tensor}
\usepackage{todonotes}

\renewcommand{\bar}[1]{\overline{#1}}

\makeatletter
\newcommand{\subalign}[1]{%
  \vcenter{%
    \Let@ \restore@math@cr \default@tag
    \baselineskip\fontdimen10 \scriptfont\tw@
    \advance\baselineskip\fontdimen12 \scriptfont\tw@
    \lineskip\thr@@\fontdimen8 \scriptfont\thr@@
    \lineskiplimit\lineskip
    \ialign{\hfil$\m@th\scriptstyle##$&$\m@th\scriptstyle{}##$\crcr
      #1\crcr
    }%
  }
}
\makeatother

\begin{document}
\pagenumbering{gobble}
\hfill
\vskip 0.01\textheight
\begin{center}
{\Large\bfseries 
Minimal models of field theories: \\ [2mm] Chiral Higher Spin Gravity}

\vspace{0.4cm}

\vskip 0.03\textheight
\renewcommand{\thefootnote}{\fnsymbol{footnote}}
Evgeny \textsc{Skvortsov}\footnote{Research Associate of the Fund for Scientific Research -- FNRS, Belgium}${}^{a,b}$ \& Richard Van Dongen${}^{a}$
\renewcommand{\thefootnote}{\arabic{footnote}}
\vskip 0.03\textheight

{\em ${}^{a}$ Service de Physique de l'Univers, Champs et Gravitation, \\ Universit\'e de Mons, 20 place du Parc, 7000 Mons, 
Belgium}\\
\vspace*{5pt}
{\em ${}^{b}$ Lebedev Institute of Physics, \\
Leninsky ave. 53, 119991 Moscow, Russia}\\

\end{center}

\vskip 0.02\textheight

\begin{abstract}
There exists a unique class of local Higher Spin Gravities with propagating massless fields in $4d$ --- Chiral Higher Spin Gravity. Originally, it was formulated in the light-cone gauge. We construct a covariant form of this theory as a Free Differential Algebra up to NLO, i.e. at the level of equations of motion. It also contains the recently discovered covariant forms of the higher spin extensions of SDYM and SDGR, as well as SDYM and SDGR themselves. From the mathematical viewpoint the result is equivalent to taking the minimal model (in the sense of $L_\infty$-algebras) of the jet-space extension of the BV-BRST formulation of Chiral Higher Spin Gravity, thereby, containing also information about (presymplectic AKSZ) action, counterterms, anomalies, etc.
\end{abstract}

\newpage
\tableofcontents
\newpage
\section{Introduction}
\label{sec:}
\pagenumbering{arabic}
\setcounter{page}{2}
Higher Spin Gravities (HiSGRA) are defined to be the smallest possible extensions of gravity with massless fields of arbitrary spin. While there are good reasons to expect higher spin states to play an important role in general, e.g. string theory, the masslessness should imitate the high energy behavior and, for that reason, HiSGRA can be interesting probes of the quantum gravity problems since some of the issues can become visible already at the classical level. Indeed, it is quite challenging to construct HiSGRA due to massless higher spin fields facing numerous issues. As a result, all concrete HiSGRA's available at the moment are quite peculiar: topological models $3d$ with (partially)-massless and conformal fields \cite{Blencowe:1988gj,Bergshoeff:1989ns,Campoleoni:2010zq,Henneaux:2010xg,Pope:1989vj,Fradkin:1989xt,Grigoriev:2019xmp}; $4d$ conformal HiSGRA  \cite{Segal:2002gd,Tseytlin:2002gz,Bekaert:2010ky} that is a higher spin extension of Weyl gravity; Chiral HiSGRA \cite{Metsaev:1991mt,Metsaev:1991nb,Ponomarev:2016lrm,Skvortsov:2018jea,Skvortsov:2020wtf}  and its truncations \cite{Ponomarev:2017nrr,Krasnov:2021nsq}.\footnote{There are also other interesting recent ideas, e.g. \cite{deMelloKoch:2018ivk,Aharony:2020omh} and \cite{Sperling:2017dts,Tran:2021ukl,Steinacker:2022jjv}.} In this paper we covariantize the interactions of Chiral HiSGRA. 

Chiral HiSGRA is easy to describe due to its simplicity --- interactions stop at the cubic level in the action. It is built from the standard cubic interactions, even though the formulation available before the present paper is in the light-cone gauge. It is advantageous that the light-cone gauge and the spinor-helicity formalism are closely related \cite{Chalmers:1998jb,Chakrabarti:2005ny,Chakrabarti:2006mb,Bengtsson:2016jfk,Ponomarev:2016cwi}. As is well-known \cite{Bengtsson:1986kh,Benincasa:2011pg}, the Lorentz invariance fixes cubic amplitudes  $V_{\lambda_1,\lambda_2,\lambda_3}$ and for any triplet of helicities $\lambda_1+\lambda_2+\lambda_3>0$ there is a unique vertex and the corresponding amplitude:
\begin{align}\label{genericV}
   V_{\lambda_1,\lambda_2,\lambda_3}\Big|_{\text{on-shell}} \sim 
    [12]^{\lambda_1+\lambda_2-\lambda_3}[23]^{\lambda_2+\lambda_3-\lambda_1}[13]^{\lambda_1+\lambda_3-\lambda_2}\,.
\end{align}
Chiral Theory can be defined as a unique combination of vertices \cite{Metsaev:1991mt,Metsaev:1991nb,Ponomarev:2016lrm} that (a) contains at least one nontrivial self-interaction of a higher spin state with itself; (b) leads to a Lorentz-invariant theory; (c) does not require higher order contact vertices. These assumptions imply that the spectrum of the theory has to contain massless fields of all spins $s=0,1,2,...$, i.e. helicities $\lambda \in(-\infty, +\infty)$ and all coupling constants are uniquely fixed to be
\begin{align}\label{eq:magicalcoupling}
    V_{\text{Chiral}}&= \sum_{\lambda_1,\lambda_2,\lambda_3}  C_{\lambda_1,\lambda_2,\lambda_3}V_{\lambda_1,\lambda_2,\lambda_3}\,, && C_{\lambda_1,\lambda_2,\lambda_3}=\frac{\kappa\,(l_p)^{\lambda_1+\lambda_2+\lambda_3-1}}{\Gamma(\lambda_1+\lambda_2+\lambda_3)}\,.
\end{align}
Here, $l_p$ is a constant of dimension length, e.g. Planck length, and $\kappa$ is an arbitrary dimensionless constant. In principle, there exists the $\phi^3$-vertex, i.e. $\lambda_i=0$, but it is not present in Chiral Theory. We also see that the $\Gamma$-function restricts the range of summation to  $\lambda_1+\lambda_2+\lambda_3>0$. All such vertices are present. For example, one has the half of the usual $+2,+2,-2$ Einstein-Hilbert vertex and, provided the Yang-Mills groups are turned one, the Yang-Mills interaction $+1,+1,-1$. Importantly, the higher derivative corrections are also needed, e.g. the half of the Goroff-Sagnotti counterterm \cite{Goroff:1985th}, which is $+2,+2,+2$. Such higher derivative terms originate from string theory as well, e.g. \cite{Metsaev:1986yb}. 

It was shown that the tree-level amplitudes vanish on-shell \cite{Skvortsov:2018jea,Skvortsov:2020wtf}. At one-loop there are no UV divergences and the one-loop amplitudes are proportional to the all helicity plus amplitudes of QCD or SDYM at one loop \cite{Skvortsov:2020gpn}. They also have a higher spin kinematical factor and a factor of the total number of degree of freedom $\sum_\lambda 1$. The latter is infinite and, as in any QFT with infinitely many fields, see e.g. \cite{Fradkin:1982kf}, has to be given a prescription for. A great deal of vacuum one-loop results \cite{Gopakumar:2011qs,Tseytlin:2013jya,Giombi:2013fka,Giombi:2014yra,Beccaria:2014jxa,Beccaria:2014xda,Beccaria:2015vaa,Gunaydin:2016amv,Bae:2016rgm,Skvortsov:2017ldz} suggest that this has to be regularized to zero.   

The power of the light-cone gauge is in that it excludes unphysical degrees of freedom and evades ambiguities of covariant (gauge) descriptions. However, many interesting questions, e.g. nontrivial backgrounds, exact solutions, higher order quantum corrections, are easier to tackle within a covariant description. Until recently a subtlety has been that Chiral Theory requires all vertices \eqref{genericV}, some of which cannot be written within the most common covariant approach to higher spin fields \cite{Conde:2016izb}, where a massless spin-$s$ field is represented by a symmetric rank-$s$ tensor $\Phi_{\mu_1...\mu_s}$. This puzzle has been resolved in \cite{Krasnov:2021nsq}, where it was shown that the most basic problematic interactions of higher spin fields --- Yang-Mills and gravitational --- can easily be constructed by employing the covariant field variables discovered first in Twistor Theory \cite{Atiyah:1979iu,Hitchin:1980hp, Eastwood:1981jy,Woodhouse:1985id}. This should not be surprising since Chiral Theory was shown to admit a formulation similar to self-dual Yang-Mills and self-dual gravity \cite{Ponomarev:2017nrr} and Twistor techniques are most natural for self-dual theories.  

In the present paper we extend these results to Chiral Theory and construct its minimal model or,  equivalently, its classical equations of motion as a Free Differential Algebra \cite{Sullivan77} to NLO. In other words, Chiral Theory can be written as a sigma-model $ d \Phi= Q(\Phi) $, where $\Phi$ are maps from $\Pi T\mathcal{M}$ (the algebra of differential forms on a manifold $\mathcal{M}$) to another supermanifold $\mathcal{N}$ equipped with a homological vector field $Q$, $QQ=0$. All essential information about a given theory, e.g. action, anomalies, etc., is encoded in its minimal model as the $Q$-cohomology \cite{Barnich:2009jy,Grigoriev:2019ojp}. Therefore, the results of the paper can be used to investigate the quantum properties of Chiral Theory, as well as to construct an action and look for classical solutions.  

The paper is organized as follows. After a brief introduction in section \bref{sec:mm} into Free Differential Algebras and minimal models, we give in section \bref{sec:initial} a concise overview of \cite{Krasnov:2021nsq} where covariant actions for the higher spin extensions of self-dual Yang-Mills and self-dual Gravity were constructed. These results give important hints on how to extend them to Chiral Theory, which these two classes are contractions of \cite{Ponomarev:2017nrr}. To find the right gauge algebra (higher spin algebra) is the first step and it was done in \cite{Krasnov:2021cva}. We then proceed in section \bref{sec:FDA} to the main part and construct $L_\infty$-structure maps/interaction vertices. We also check that some three-point amplitudes \eqref{eq:magicalcoupling} are correctly reproduced. The latter means that the FDA incorporates all the physically relevant information at NLO. There are still some higher structure maps to be found that are required for the complete covariantization of Chiral Theory. We leave this problem to the future work.

\section{Minimal Models}
\label{sec:mm}
There is a very useful $L_\infty$-algebra, better say a $Q$-manifold, that can naturally be associated to any (gauge) theory and encodes all relevant information about it, which is called the minimal model. As is explained in \cite{Brandt:1997iu,Brandt:1996mh,Barnich:1994db,Barnich:1994mt,Barnich:2010sw,Grigoriev:2012xg,Grigoriev:2019ojp,Grigoriev:2020lzu}, one begins with the jet space BV-BRST formulation of a given (gauge) theory. This way one gets a huge $L_\infty$-algebra which has been quite useful in the analysis of numerous problems in (quantum) field theories, see e.g. \cite{Barnich:1994db,Barnich:1994mt}. One can then consider various equivalent reductions of this algebra that are quasi-isomorphic to it. An important step is to take a usually much smaller equivalent $L_\infty$-algebra, called its minimal model.  The minimal model is, in some sense, the smallest possible $L_\infty$-algebra associated to a given field theory. Nevertheless, modulo the usual topological issues, it contains the full information about invariants, conserved currents, actions, counterterms, anomalies, etc. of the initial field theory \cite{Barnich:2009jy,Grigoriev:2019ojp}. 

Given a BRST complex that is non-negatively graded, e.g. the minimal model, one can consider an associated sigma model whose fields are coordinates on the above $Q$-manifold \cite{Barnich:2010sw}:  
\begin{align}\label{mevenmore}
    d \Phi&= Q(\Phi) \,.
\end{align}
Here, $\Phi\equiv \Phi(x,dx)$ are maps $\Pi T\mathcal{M} \rightarrow \mathcal{N}$ from the exterior algebra of differential forms on a space-time manifold $\mathcal{M}$ to a supermanifold $\mathcal{N}$ that is equipped with a homological vector field $Q$, $QQ=0$. Equations \eqref{mevenmore} and their natural gauge symmetries are equivalent to the initial field theory,\footnote{In general, the equations describe the parameterized version of the initial gauge field theory \cite{Barnich:2010sw}.} thereby providing its reformulation as a Free Differential Algebra.\footnote{Sullivan introduced Free Differential Algebras in \cite{Sullivan77} together with minimal models in the case of differential graded Lie algebras. FDA were re-introduced into physics \cite{vanNieuwenhuizen:1982zf,DAuria:1980cmy} in the supergravity context and a bit later in the higher spin gravity context in \cite{Vasiliev:1988sa}. }

If $\Phi^\aA$ are coordinates on $\mathcal{N}$, $QQ=0$ is equivalent to \eqref{mevenmore} being formally consistent (that is $dd=0$ does not lead to any algebraic constraints on the fields), which can be rewritten as
\begin{align}
    Q^2&=0 &&\Longleftrightarrow && Q^\aB \frac{\pl}{\pl \Phi^\aB} Q^\aA=0\,.
\end{align}
The latter condition, when Taylor expanded in $\Phi$, is equivalent to the $L_\infty$-relations \cite{Stasheff,Alexandrov:1995kv} that define an $L_\infty$-algebra. This shows that FDA, $L_\infty$ and $Q$-manifolds are all closely related.  
In many practical applications of minimal models, e.g. gauge field theories including gravity,\footnote{For some of the supergravities forms of higher degree need to be introduced.}  coordinates on the formal graded manifold $\mathcal{N}$ consist of two subsets: degree-one and degree-zero. We denote the coordinates and, then, the corresponding fields $\omega$ and $C$, respectively. From the space-time point of view $\omega$ becomes a one-form connection of some Lie algebra and zero-form $C$ becomes a matter field taking values in some representation $\rho$. The simplest system one can write
\begin{align}
    d\omega &=\tfrac12[\omega,\omega]\,,& dC&=\rho(\omega)C \label{laxd}\,,
\end{align}
consists of the flatness condition for $\omega$ and of the covariant constancy equation on $C$. These two equations will describe a background and the physical degrees of freedom propagating on it. The most general non-linear deformation reads\footnote{It was first proposed in \cite{Vasiliev:1988sa} to look for Higher Spin Gravities in the form of an FDA. However, it is important to constrain the vertices by further conditions: (a) to restrict to a basis of independent interaction vertices (otherwise one and the same interaction can be present in infinitely many equivalent but differently looking forms); (b) to impose some form of locality (otherwise any deformation can be completed with higher orders \cite{Barnich:1993vg}, or, in the light-cone gauge, any function can serve as a Hamiltonian unless we care about locality of the boost generators). All these issues are present \cite{Boulanger:2015ova,Skvortsov:2015lja} in \cite{Vasiliev:1988sa}. Therefore, unless (a) and (b) are taken into account $Q$ just gives the most general ansatz for interactions consistent with symmetries rather than any concrete theory. These issues are under control in the present paper.  }
\begin{equation}\label{mostgeneral}
    \begin{array}{rcl}
         d\omega&=&l_2(\omega,\omega)+l_3(\omega,\omega,C)+l_4(\omega,\omega,C,C)+\ldots\,,\\
    dC&=&l_2(\omega,C)+l_3(\omega,C,C)+\ldots\,.
    \end{array}
\end{equation}
This algebraic structure can also be identified as a Lie algebroid. Here the initial data ---  Lie algebra and its module --- are encoded in the bilinear maps $l_2(\omega,\omega)$ and $l_2(\omega,C)$, respectively. The higher spin algebra for Chiral Theory was guessed in \cite{Krasnov:2021cva} based on its truncation to the self-dual gravity sector. The module structure is easy to identify, see below. The problem is to find the higher order vertices. In the paper we determine $l_3(\bullet,\bullet,\bullet)$.

\section{HS-SDYM and HS-SDGR}
\label{sec:initial}
Since the full covariant form of Chiral HiSGRA is not known and this is exactly the problem we address in the paper, a good starting point is to extract some useful information from the two contractions of Chiral Theory \cite{Ponomarev:2017nrr,Krasnov:2021nsq}, which can be understood as higher spin extensions of SDYM and SDGR. We begin by reviewing some necessary facts about free fields. Impatient readers familiar with the formalism can skip to section \bref{sec:FDA}.

\subsection{Free fields}
\label{sec:free}
Free massless fields of any spin can be described by equations proposed by Penrose \cite{Penrose:1965am}\footnote{We also introduce a compact notation for symmetric indices: all indices in which some tensor is symmetric or to be symmetrized are denoted by the same letter. In addition a group of $k$ symmetric indices $A_1...A_k$ can be abbreviated as $A(k)$. }
\begin{align}\label{hsA}
    \nabla^\fdu{B}{A'} \Psi^{BA(2s-1)}&=0\,, &\nabla\fud{A}{B'} \Psi^{B'A'(2s-1)}&=0\,.
\end{align}
The equations help to separate helicity eigenstates: one of them describes, say positive, and another the negative helicity states. Twistor theory is very handy in constructing self-dual theories. It requires positive and negative helicity states be described asymmetrically \cite{Hughston:1979tq,Eastwood:1981jy,Woodhouse:1985id}
\begin{align}\label{hsB}
    \nabla\fdu{A}{A'}\Phi^{A,A'(2s-1)}&=0\,, && \delta \Phi^{A,A'(2s-1)}=\nabla^{A A'}\xi^{A'(2s-2)}\,,
\end{align}
where $\Phi^{A_1...A_{2s-1},A'}$ is a gauge potential. For $s=1$ it coincides with the usual one $A_\mu \sim \Phi^{A,A'}$. For $s=2$ it can be identified with a component of the spin-connection. A bit more geometrically one can \cite{Hitchin:1980hp} introduce a one-form connection 
\be
\omega^{A'(2s-2)}=\Phi^{B,A'(2s-2)'B'} dx_{BB'}\,.
\ee
It can be decomposed into two irreducible spin-tensors
\begin{align}
    \omega^{A'(2s-2)}\equiv e_{BB'}\Phi^{B,A'(2s-2)B'}+e\fdu{B}{A'}\Theta^{B,A'(2s-3)}\,,
\end{align}
where $e^{AA'}\equiv e^{AA'}_\mm \, dx^\mm$ is the vierbein one-form. With the help of gauge transformations
\begin{align}\label{lin-gauge}
    \delta \omega^{A'(2s-2)}&= \nabla \xi^{A'(2s-2)} +e\fdu{C}{A'} \eta^{C,A'(2s-3)}\,,
\end{align}
we get \eqref{hsB} for $\Phi$ and can eliminate $\Theta$. Eqs. \eqref{hsA} and \eqref{hsB} follow from a simple action \cite{Hitchin:1980hp,Krasnov:2021nsq}\footnote{This action also can be derived as the presymplectic AKSZ action \cite{Sharapov:2021drr}.} 
\begin{align}\label{niceaction}
    S= \int \Psi^{A'(2s)}\wedge H_{A'A'}\wedge \nabla \omega_{A'(2s-2)}\,.
\end{align}
Here $H^{A'B'}\equiv e\fdu{C}{A'}\wedge e^{CB'}$. For $s=1$ we have the action of the free SDYM theory. By replacing $\nabla \omega$ with $F=\nabla \omega-\tfrac12[\omega, \omega]$ and promoting $\omega$ and $\Psi$ to a Lie-algebra-valued one-form we get the complete SDYM action \cite{Krasnov:2021nsq}.

\paragraph{Free equations of motion as Free Differential Algebra.} Let us start\footnote{The content of this paragraph has a large overlap with original paper \cite{Vasiliev:1986td}. Apart from the self-dual subtleties the material is standard and can be found, e.g., in \cite{Didenko:2014dwa}.} with the variational equations of motion, which do not have an FDA-form yet:
\begin{align}\label{first}
    \nabla \Psi^{A'(2s)}\wedge H_{A'A'}&=0\,, && H^{A'A'}\wedge \nabla \omega^{A'(2s-2)}=0\,.
\end{align}
Indeed, we need $\nabla \Psi=...$ and $\nabla \omega=...$. The equations are equivalent to
\begin{align}
    \nabla \Psi^{A'(2s)}&= e_{BB'}\Psi^{B,A'(2s)B'}\,,
&
    \nabla \omega^{A'(2s-2)} &= e\fdu{B}{A'} \omega^{B,A(2s-3)}\,,
\end{align}
where we introduced a zero-form $\Psi^{A,A'(2s+1)}$ and one-form $\omega^{A,A'(2s-3)}$. These fields are known to be relevant for free higher spin fields since \cite{Vasiliev:1986td}.\footnote{Indeed, since \cite{Vasiliev:1986td} introduces fields to parameterize all on-shell nontrivial derivatives of massless fields, any other covariant formulation has to employ at least some of them. Note, however, that the fields of \eqref{first} appeared first thanks to the twistor approach \cite{Penrose:1965am,Hughston:1979tq,Eastwood:1981jy,Woodhouse:1985id}.} 
Of course, we need to know what $\nabla$ of these new fields is, which encourages one to introduce other fields and so on. It is clear that the free equations are easy to write as
\besubeqs
\begin{align}
    d\omega^{A(i),A'(n-i)}&= e\fdu{B}{A'} \omega^{A(i)B,A'(n-i-1)}\,, && i=0,...,n-1\,,\\
    d\omega^{A(n)}&= H_{BB} C^{A(n)BB}\,,\\
    dC^{A(n+k+2),A'(k)}&= e_{BB'} C^{BA(n+k+2),B'A'(k)}\,, && k=0,1,2,...\,,\\
    d\Psi^{A(k),A'(n+k+2)}&= e_{BB'} \Psi^{BA(k),A'(n+k+2)B'}\,, && k=0,1,2,...\,,
\end{align}
\esubeqs
where $C$ and $\Psi$ are zero forms and $\omega$ are one-forms. It is convenient introduce generating functions:
\begin{align}
    \omega(y,\bry)&= \sum_{n,m}\tfrac{1}{n!m!} \omega_{A(n),A'(m)}\, y^A...y^A\, \bry^{A'}...\bry^{A'}\,,  
\end{align}
\textit{idem}. for $C$, where we pack both $C^{A(k),A'(n+k+2)}$ and $\Psi^{A(n+k+2),A'(k)}$ into a single generating function $C(y,\bry)$. On top of that $C(y,\bry)$ contains $C^{A(k),A'(k)}$, which describe a free massless scalar field. Note that the scalar field is necessarily present in Chiral Theory. We can summarize the free equations as (recall that $\nabla^2=0$)
\begin{align}\label{linearizeddata}
    \nabla\omega &= e^{BB'}y_{B'} \pl_{B} \omega +H^{BB} \pl_{B}\pl_{B}C(y,\bry=0)\,,& 
    \nabla C&= e^{BB'}\pl_B \pl_{B'} C\,.
\end{align}
These equations form a boundary condition for the non-linear theory.  

\subsection{Initial data for interactions}
\label{sec:}
It can be useful to have a look at the two contractions of Chiral Theory \cite{Ponomarev:2017nrr,Krasnov:2021nsq} in order to understand how interactions can be introduced. Both HS-SDYM and HS-SDGR \cite{Krasnov:2021nsq} operate with holomorphic fields $\omega^{A'(2s-2)}$ and $\Psi^{A'(2s)}$. It is still useful to package them into generating functions $\omega(\bry)$ and $\Psi(\bry)$.

\paragraph{HS-SDYM.} In order to construct Yang-Mills type interactions of higher spin fields, we promote $\omega$ and $\Psi$ to Lie-algebra-valued fields, e.g. $\omega^{A'(k)}\equiv \omega^{A'(k);a}\, T_a$. It is convenient to realize $T_a$ as matrices $\mathrm{Mat}_N$ for some $N$, e.g. $\omega(\bry)\equiv \omega(\bry)\fud{i}{j}$. We will omit the matrix indices and the only trace they leave is that we cannot swap various $\omega$ and $\Psi$ factors, the order is important, e.g. $\Psi\wedge \omega\neq \omega\wedge\Psi$. The action of HS-SDYM can be written as
\begin{align}\label{HSSDDYM}
    S&=\sum_{s=1}\tfrac{1}{(2s)!}\tr\int \Psi^{A'(2s)} \wedge H_{A'A'}\wedge F_{A'(2s-2)}\,,
\end{align}
where the curvature is $F(\bry)=\nabla\omega-\omega\wedge \omega$. Note that indices contracted with $\bry^{A'}$ are symmetrized automatically:
\begin{align}\label{commutatorYM}
    \omega \wedge \omega &= \sum_{n,m=0} \frac{1}{2\,n!m!}\,[\omega_{A'(n)}, \omega_{A'(m)}]\, \bry^{A'_1}...\,\bry^{A'_{n+m}}\,.
\end{align}
The action is invariant under the Yang-Mills transformations:
\begin{align}
    \delta \omega&= \nabla\xi-[\omega,\xi]\,,& 
    \delta \Psi&= [\Psi,\xi]\,.
\end{align}
It is also invariant under the algebraic symmetries (thanks to $e^{BA'}\wedge H^{A'A'}\equiv0$):
\begin{align}\label{symmetry-shifts}
    \delta \omega^{A'(k)}&= e\fdu{C}{A'} \eta^{C,A'(k-1)}\,,
\end{align}
which is vital for $\omega$ to have the right number of degrees of freedom. See \cite{Krasnov:2021nsq} for detail and \cite{Tran:2021ukl} for the twistor reformulation.

In principle, we can write down the variational equations of motion and try to represent them as an FDA. Two important hints will play a role in what follows: (a) interactions must contain \eqref{commutatorYM}, i.e. $d\omega(\bry)=\omega(\bry)\wedge \omega(\bry)+...$; (b) $\Psi(\bry)$ takes values in the module that is dual to that of $\omega$, which follows from the structure of the action.  

\paragraph{HS-SDGR.} Higher spin extension of SDGR \cite{Krasnov:2016emc} is more peculiar \cite{Krasnov:2021nsq}. Let us start with its version on constant (non-zero) curvature spacetimes. The flat-space version \cite{Krasnov:2021cva} is a simple limit. To proceed we introduce a Poisson structure on the space $\mathbb{C}[\bry]$ of functions in $\bry^{A'}$:\footnote{This algebra is also know as $w_{1+\infty}$, see e.g. \cite{Adamo:2021lrv} for the latest applications.}
\begin{align}
    \{f,g\}&= \pl^{C'} f \pl_{C'}g= \sum_{n,m} \tfrac{1}{(n-1)!(m-1)!} f\fdu{A'(n-1)}{C'}g_{A'(m-1)C'}\, \bry^{A'}...\,\bry^{A'}\,.
\end{align}
Since Poisson implies Lie, we can define a curvature as usual
\begin{align}
    F&= d\omega-\tfrac12\{\omega,\omega \}\,, & \delta \omega&=d\xi -\{\omega,\xi\}\equiv D\xi\,.
\end{align}
In particular, the Poisson bracket reproduces the standard $F^{AB}= d\omega^{AB} +\omega\fud{A}{C} \wedge\omega^{CB}$ in the spin-two sector. The action reads:
\begin{align}\label{HSSDGRAads}
   S&=\tfrac12\action{\Psi}{F\wedge F} = \sum_{n,m=0}\frac{1}{2(n+m)!}\int \Psi^{A'(n+m)}\wedge F_{A'(n)}\wedge F_{A'(m)}\,.
\end{align}
It is again important that there is a generalization of the shift symmetry that leaves the full action invariant \cite{Krasnov:2021nsq}. To this effect, one first needs to induce the module structure on $\Psi$, which is a module dual to the Poisson algebra as a Lie algebra:
\begin{align}
     \scalar{ f}{ \{\xi,g\}}&:=  \scalar{ f\circ\xi}{g}\,.
\end{align}
That it is a module structure is manifested by 
\begin{align}
    \mathcal{R}_f (\Psi)&:=-\Psi\circ f\,, && [\mathcal{R}_f,\mathcal{R}_g] (\Psi)= \mathcal{R}_{\{f,g\}} (\Psi) \,.
\end{align}
The structure of the action and of the gauge symmetries gives a strong support to the idea that $\Psi$ has to be in the dual (coadjoint) representation of the higher spin symmetry. The flat-space limit is easy to take: one just needs to drop $\{\omega,\omega \}$-term in the curvature, which is equivalent to taking the commutative limit for $\bry$.  While we could discuss the FDA formulation of this theory, an example of SDGR gives enough information about the gauge algebra to attack the main problem.

\paragraph{SDGR in flat space.} It may be useful to recall the first few terms of the FDA for self-dual gravity \cite{Siegel:1992wd,AbouZeid:2005dg} in flat space \cite{SDFDA}. The action reads \cite{Krasnov:2021cva}
\begin{align}
    \int \Psi^{A'B'C'D'}\wedge d\omega_{A'B'} \wedge d\omega_{C'D'}\,.
\end{align} 
The equations of motion are ($F^{A'B'}\equiv d\omega^{A'B'}$)
\begin{align}
    F_{(A'B'} \wedge F_{C'D')}&=0\,, & d\Psi^{A'B'C'D'}\wedge F_{A'B'}&=0\,.
\end{align}
The first equation implies that there is no $5$-dimensional representation of $sl_2$ in the symmetric tensor product of two $F^{A'B'}$. Therefore, $F^{A'B'}$ can be represented as $e\fdu{B}{A'}\wedge e^{BA'}$ for some field $e^{AA'}$. Indeed, it is easy to see that $F^{A'A'}\wedge F^{A'A'}=0$. Now, it is not surprising that the first few equations in the FDA read
\begin{align*}
d \omega^{A'A'}&= e\fdu{B}{A'}\wedge e^{BA'}\,,
    & d e^{AA'}&=\omega\fud{A}{B}\wedge e^{BA'} \,,
   & d \omega^{AA}&=\omega\fud{A}{C}\wedge\omega^{CA}+H_{BB}C^{AABB}\,.
\end{align*}
We note that the non-abelian terms with $\omega^{A'A'}$ are missing here-above as compared to the standard curvature of $so(3,2)\sim sp(4)$. However, we do not recognize the curvature of the Poincare algebra either. As for $\Psi$, the equation can be rewritten as 
\begin{align}
    d\Psi^{A'B'C'D'}\wedge H_{A'B'}&=0\,,
\end{align}
which is equivalent to
\begin{align}
    \nabla\Psi^{A'A'A'A'}&= e_{BB'}\Psi^{B,A'A'A'A'B'}\,.
\end{align}
One can see that we employ exactly the same fields as for the full gravity, but certain structures 'abelianize'. Half of the Lorentz symmetry becomes global rather than originating from a local gauge symmetry. 

\section{FDA for Chiral Higher Spin Gravity}
\label{sec:FDA}
After the preliminary steps above we proceed to constructing the Free Differential Algebra of Chiral Theory. Firstly, we summarize the known initial data and boundary conditions for the $L_\infty$ structure maps. 

\subsection{Initial data}
\label{sec:}
\paragraph{Coordinates/fields, on-shell jet.} The coordinates on the $Q$-manifold or, alternatively, the fields of the minimal model are exactly the same as for the free fields discussed in Section \ref{sec:initial}. 
\besubeqs\label{spec}
\begin{align}
    h&=+s: && \omega^{A(k),A'(2s-2-k)}\,, C^{A(2s+i),A'(i)} \,, \qquad k=0,...,2s-2\,,\quad i=0,1,2,... \,,\\
    h&=-s: && C^{A(i),A'(2s+i)}\,,\qquad i=0,1,2,... \,,\\
    h&=0: && C^{A(i),A'(i)}\,,\qquad i=0,1,2,...\,.
\end{align}
\esubeqs
As before, it is convenient to keep all components of $\omega$ and $C$ confined in generating functions $\omega(y,\bry)$, $C(y,\bry)$. Chiral Theory is known to admit Yang-Mills gaugings \cite{Skvortsov:2020wtf} that, however, come in a very restricted Chan-Paton-like fashion. To be precise, one can have $U(N)$, $O(N)$ and $USp(N)$ gaugings. Therefore, we assume that $\omega$ and $C$ take values in $\mathrm{Mat}_N$.\footnote{It was shown in \cite{Sharapov:2018kjz,Sharapov:2019vyd,Sharapov:2020quq} that this assumption allows one to reduce a complicated Chevalley-Eilenberg cohomology problem to a much simpler Hochschild one. In other words, it is important to remember that usually higher spin algebras originate from associative ones. }

\paragraph{General form.} Given all the data above, we are looking for Chiral Theory in the form
\besubeqs\label{eq:chiraltheory}
\begin{align} 
    d\omega&= \mathcal{V}(\omega, \omega) +\mathcal{V}(\omega,\omega,C)+...\,,\\
    dC&= \mathcal{U}(\omega,C)+ \mathcal{U}(\omega,C,C)+... \,.
\end{align}
\esubeqs
Here, $\mathcal{V}$ and $\mathcal{U}$ are some $L_\infty$ structure maps to be determined. It would be sufficient if the expansion stops at the quartic terms. This can be justified on the basis of the light-cone action of Chiral Theory: interactions stop at the cubic level. One might argue that they have to stop then at quadratic terms for equations. However, this does not have to be the case since the light-cone gauge theory requires a background, i.e. some specific $\omega_0$. Therefore, $\mathcal{V}(\omega,\omega,C)$ is legit, as well as $\mathcal{V}(\omega,\omega,C,C)$, while higher order terms may not be necessary. One can also see that $\mathcal{V}(\omega, \omega) $ cannot account for all of the interactions, e.g. $\omega$ does not contain the scalar field at all. 

An important subtlety is that covariantization of a given theory (going from the light-cone gauge to a covariant formulation) may require more terms in the perturbation theory that are there only for the sake of covariance. Such contact terms will not give any contribution to physical amplitudes. Another subtlety is due to field redefinitions: it is easy to perform a nonlinear field redefinition in the cubic theory and generate spurious interactions. Alternatively, when looking for $\mathcal{V}$'s and $\mathcal{U}$'s one can find oneself in an unfortunate field frame with such spurious interactions all around. We check in Appendix \ref{app:} that certain cubic amplitudes are reproduced correctly. Therefore, \eqref{eq:chiraltheory} contains all the physically relevant information. Comparing the Chiral Theory FDA to those of SDYM and SDGR \cite{SDFDA} we find that the former contains only the terms essential for consistency, which fixes field redefinitions. 

\paragraph{Boundary conditions.} There are some boundary conditions for $\mathcal{V}$'s and $\mathcal{U}$'s that we learned from the free equations \eqref{linearizeddata}:
\besubeqs\label{eq:boundaryconditions}
\begin{align}
    \mathcal{V}(e,\omega)+\mathcal{V}(\omega,e)&=e^{CC'} \pl_C \bry_{C'} \omega\,,\\
    \mathcal{U}(e,C)+\mathcal{U}(C,e)&=e^{CC'} \pl_C \pl_{C'} C\,,\\
    \mathcal{V}(e,e,C)&= e\fud{C}{B'}e^{CB'} \pl_C \pl_C C(y,\bry=0)\,.
\end{align}
\esubeqs
To summarize we are looking for a theory with the spectrum of fields given in \eqref{spec}, in the form of FDA \eqref{eq:chiraltheory} such that it reproduces the boundary conditions \eqref{eq:boundaryconditions}, i.e. the free equations. 

\subsection{FDA}
\label{sec:}
In what follows we will have to write down ans{\"a}tze for $L_\infty$-maps. Given that we have packaged the coordinates into generating functions $\omega(y,\bry)$ and $C(y,\bry)$, the $L_\infty$-structure maps can be represented by poly-differential operators:
\begin{align}
    \mathcal{V}(f_1,...,f_n)&= \mathcal{V}(y, \pl_1,...,\pl_2)\, f(y_1)...f(y_n) \Big|_{y_i=0}\,,
\end{align}
where $f_i$'s are $\omega$'s or $C$'s and we have explicitly indicated dependence on $y$, omitting $\bry$ which can be treated similarly. With further details on the operator calculus collected in Appendix \ref{app:}, we only note that (i) we abbreviate $\bry^{A'}\equiv p_0^{A'}$, $\pl^{\bry_i}_{A'}\equiv p_{A'}^i$, $y^A\equiv q_0^A$, $\pl^{y_i}_A\equiv q_A^i$; (ii) contractions $p_{ij}\equiv p_i \cdot p_j\equiv -\epsilon_{AB}p^A_{i}p_{j}^B=p^A_{i}p_{jA}$ are done in such a way that $\exp[p_0\cdot p_i]f(y_i)=f(y_i+y)$; (iii) all operators are Lorentz invariant in the most naive sense of having all indices contracted either with $\epsilon_{AB}$ or $\epsilon_{A'B'}$; (iv) we usually omit explicit arguments $y_i$ in $f$'s, drop $|_{y_i=0}$ and sometimes write down only the operator itself whenever it is clear what the arguments are. Of course, all poly-differential operators are assumed to be local, i.e. they map polynomials to polynomials, which, after Taylor expansion means, that the operators contract a number of Lorentz indices on the arguments.\footnote{Note that this locality is just a requirement for $\mathcal{V}$ to imply some contraction of Lorentz indices (hidden by $y$) on the arguments, which is a type of locality used in \cite{Vasiliev:1988sa}. The locality in the field theory sense is more subtle --- one has to control the number of derivatives in interactions. The interactions in the present paper are local as in Chiral Theory, i.e. vertices contain a finite number of derivatives provided the helicities of the fields at a given vertex are fixed. } To give a couple of useful examples, the usual commutative product on $\bry$ and the Moyal-Weyl star-product on $y$ correspond to the following symbols
\besubeqs
\begin{align}
    \exp{(\bar{y}(\bar{\partial}_1+\bar{\partial}_2))}&= \exp[p_0 \cdot p_1+p_0 \cdot p_2]\equiv \exp[p_{01}+p_{02}]\,, \\
    \exp{(y(\partial_1+\partial_2)+\partial_1\partial_2})&= \exp[q_0 \cdot q_1+q_0 \cdot q_2+q_1 \cdot q_2]\equiv \exp[q_{01}+q_{02}+q_{12}]\,.
\end{align}
\esubeqs
We also would like to rewrite  the boundary conditions \eqref{eq:boundaryconditions} in the operator language:
\besubeqs\label{eq:boundaryconditionsB}
\begin{align}
    \mathcal{V}(e,\omega)+\mathcal{V}(\omega,e)&\sim p_{01}q_{12}\, e^{p_{02}+q_{02}}\, ( e^{CC'}y^1_C \bry^1_{C'})\, \omega(y_2,\bry_2)\Big|_{y_{1,2}=\bry_{1,2}=0}\label{eq:boundaryconditionsBA}\,,\\
    \mathcal{U}(e,C)+\mathcal{U}(C,e)&\sim q_{12} p_{12}\, e^{p_{02}+q_{02}}\, ( e^{CC'}y^1_C \bry^1_{C'})\, C(y_2,\bry_2)\Big|_{y_{1,2}=\bry_{1,2}=0}\label{eq:boundaryconditionsBB}\,,\\
    \mathcal{V}(e,e,C)&\sim  q_{13} q_{23} p_{12}\, e^{q_{03}}\,( e^{BB'}y^1_B \bry^1_{B'})( e^{CC'}y^2_C \bry^2_{C'})\, C(y_3,\bry_3)\Big|_{y_{1,2,3}=\bry_{1,2,3}=0}\label{eq:boundaryconditionsBC}\,,
\end{align}
\esubeqs
where the $\sim$ sign means that in the actual FDA we only care about reproducing these structures up to an overall coefficient. The last boundary condition, if satisfied, ensures the nontriviality of the full vertex. We will also give a rigorous proof of this fact. 

\paragraph{Higher spin algebra.} The $L_\infty$-relations or the formal consistency of \eqref{eq:chiraltheory} at order $\omega^3$ imply the Jacobi identity for $\mathcal{V}(\bullet,\bullet)$
\begin{equation}\label{eq:omegacubed}
    \mathcal{V}(\mathcal{V}(\omega,\omega),\omega)-\mathcal{V}(\omega,\mathcal{V}(\omega,\omega))=0\,.
\end{equation}
The presence of the matrix factors reduces the Jacobi identity to a much simpler and more restrictive associativity condition, i.e $\mathcal{V}(a,b)$ must define an associative product, where $a,b\in \mathbb{C}[y,\bry]$. Given the nonlinear pieces of various (sub)theories there are not so many associative algebras one can think of. In fact, the only option \cite{Krasnov:2021cva} is to define\footnote{A very similar algebra in the same context, but in the light-cone gauge, appeared even before \cite{Ponomarev:2017nrr}. }
\begin{align}
    \mathcal{V}(f,g)&=c\exp{[q_{01}+q_{02}+q_{12}]} \exp{[p_{01}+p_{02}]}f(y_1,\bar{y}_1)\wedge g(y_2,\bar{y}_2)\Big|_{y_i=\bar{y}_i=0}\equiv f\star g\,,
\end{align}
with $c$ an undetermined prefactor. In words $\mathcal{V}(f,g)\equiv f\star g$ is the commutative product on $\bry$ and the star-product on $y$. Therefore, as the higher spin algebra $\hs$ we take the tensor product of the Weyl algebra in $y$ and of the commutative algebra of function in $\bry$, $\hs = A_1\otimes \mathbb{C}[\bry]$. In addition we assume the matrix factor $\mathrm{Mat}_N$. This choice for $\mathcal{V}(\omega,\omega)$ is also consistent with the boundary conditions in equation \ref{eq:boundaryconditionsBA}:
\begin{equation}
    \mathcal{V}(e,\omega)+\mathcal{V}(\omega,e)=2c\, e^{BB'}\bar{y}_B'\partial_B\omega\,,
\end{equation}
which encourages us to set $c=\frac{1}{2}$, so that
\begin{empheq}[box=\fbox]{align} \label{V(w,w)}
     \mathcal{V}(f,g)&=\tfrac12 \exp{[q_{01}+q_{02}+q_{12}]} \exp{[p_{01}+p_{02}]}
\end{empheq}

\paragraph{Coadjoint module.} Similarly, the formal consistency implies that $U(\bullet,\bullet)$ defines a representation of the higher spin algebra:
\begin{equation}
    \mathcal{U}(\mathcal{V}(\omega,\omega),C)-\mathcal{U}(\omega,\mathcal{U}(\omega,C))=0\,.
\end{equation}
The actions of HS-SDYM and HS-SDGR strongly suggest that $C(0,\bry)$ lives in the space dual to $\omega(0,\bry)$. The action on the dual space (dual to the commutative algebra of functions in $\bry$) can be defined via $\bry_A \rightarrow \alpha \pl_{A'}$, where $\alpha$ is any number. In other words, the commutative algebra of functions in $\bry$ acts on the dual space via differential operators.\footnote{Since understanding that $C$ lives in the dual module has been important for the present paper and this idea is slightly different from the folklore in the literature, we elaborate on it more in Appendix \ref{app:coadjoint}. } In terms of symbols of operators we can write
\begin{align}
    \omega(f)&= \exp{[p_{02}+\alpha p_{12}]}\, \omega(\bar{y}_1) f(\bar{y}_2)\Big|_{\bar{y}_i=0}\,.
\end{align}
With indices explicit we find
\begin{align}
    \omega(f)&=\sum_{i,n} \tfrac{\alpha^i}{n!}\, \omega^{B'(i) }f_{A'(n) B'(i)}\, \bry^{A'}...\bry^{A'}\,.
\end{align}
It is plausible to extend the idea with the dual space to the complete space $\mathbb{C}(y,\bry)$, it is unclear how to induce the module structure, otherwise. Now it is time to remember about the matrix factors. We consider functionals based on their ordering of $\omega$ and $C$:
\begin{align}
    \mathcal{U}(\omega,C)&=\mathcal{U}_1(\omega,C)+\mathcal{U}_2(C,\omega)\,.
\end{align}
The consistency condition splits into the following equations.
\begin{equation}
    \begin{split}
        &\mathcal{U}_1(\mathcal{V}(\omega,\omega),C)-\mathcal{U}_1(\omega,\mathcal{U}_1(\omega,C))=0\,,\\
        &\mathcal{U}_2(\mathcal{U}_1(\omega,C),\omega)-\mathcal{U}_1(\omega,\mathcal{U}_2(C,\omega))=0\,,\\
        &\mathcal{U}_2(\mathcal{U}_2(C,\omega),\omega)+\mathcal{U}_2(C,\mathcal{V}(\omega,\omega))=0\,.
    \end{split}
\end{equation}
In words, we have a right and a left actions of the higher spin algebra on $\mathbb{C}(y,\bry)$. The actions must be compatible with each other, which is the middle equation. Given that $C$ should be in the dual module, the structure maps $\mathcal{U}_{1,2}$ are easy to fix to be:
\begin{equation}\boxed{
    \begin{split}
        &\mathcal{U}_1(\omega,C)=+\tfrac{1}{2}\exp{[q_{01}+q_{02}+q_{12}]}\exp{[p_{02}+p_{12}]}\, \omega(y_1,\bar{y}_1) C(y_2,\bar{y}_2)\Big|_{y_i=\bar{y}_i=0}\\
        &\mathcal{U}_2(C,\omega)=-\tfrac{1}{2}\exp{[q_{01}+q_{02}+q_{12}]}\exp{[p_{01}-p_{12}]}\, C(y_1,\bar{y}_1) \omega(y_2,\bar{y}_2)\Big|_{y_i=\bar{y}_i=0}
    \end{split}}
\end{equation}
It is easy to check that boundary condition \eqref{eq:boundaryconditionsBB} is satisfied with coefficient $1$.

\paragraph{Cubic Vertex $\boldsymbol{\mathcal{V}(\omega,\omega,C)}$.} As a next step we turn to the cocycle $\mathcal{V}(\omega,\omega,C)$. It has a right to be called a cocycle. Indeed, the bilinear structure maps of any FDA (more generally, of any $L_\infty$-algebra) define a graded Lie algebra. Let us pack them into $Q_0$, $(Q_0)^2=0$. Next we look for the first order deformation $Q_1$ of $Q_0$. It is clear that $Q_1$ must be in the cohomology of $Q_0$. The action of $Q_0$ on $Q_1$ is that of the Chevalley-Eilenberg differential, according to which ${\mathcal{V}(\omega,\omega,C)}$ is a two-cocycle with values in $\hs\otimes \hs$: it takes values in $\hs$ and $C$ is in $\hs^*$. To find the equation for ${\mathcal{V}(\omega,\omega,C)}$ we evaluate the $\omega^3C$ terms after applying $d$ to \ref{eq:chiraltheory}, which leads to
\begin{align}\notag
    \mathcal{V}(\mathcal{V}(\omega,\omega,C),\omega)-\mathcal{V}(\omega,\mathcal{V}(\omega,\omega,C))&+\mathcal{V}(\mathcal{V}(\omega,\omega),\omega,C)\\
    &-\mathcal{V}(\omega,\mathcal{V}(\omega,\omega),C)+\mathcal{V}(\omega,\omega,\mathcal{U}(\omega,C))=0\,.
\end{align}
Like we did for $\mathcal{U}(\omega,C)$, we can split $\mathcal{V}(\omega,\omega,C)$ into three vertices, with different ordering of $\omega$ and $C$:
\begin{align}
    \mathcal{V}(\omega,\omega,C)=\mathcal{V}_1(\omega,\omega,C)+\mathcal{V}_2(\omega,C,\omega)+\mathcal{V}_3(C,\omega,\omega)    \,.
\end{align}
The consistency condition should now be evaluated for each ordering of $\omega$ and $C$ separately, which leads to
\begin{equation}
    \begin{split}
        &\mathcal{V}_1(\mathcal{V}(\omega,\omega),\omega,C)-\mathcal{V}(\omega,\mathcal{V}_1(\omega,\omega,C))+\mathcal{V}_1(\omega,\omega,\mathcal{U}_1(\omega,C))-\mathcal{V}_1(\omega,\mathcal{V}(\omega,\omega),C)=0\,,\\
        &\mathcal{V}(\mathcal{V}_1(\omega,\omega,C),\omega)+\mathcal{V}_1(\omega,\omega,\mathcal{U}_2(C,\omega))+\mathcal{V}_2(\mathcal{V}(\omega,\omega),C,\omega)-\mathcal{V}(\omega,\mathcal{V}_2(\omega,C,\omega))\\
        &-\mathcal{V}_2(\omega,\mathcal{U}_1(\omega,C),\omega)=0\,,\\
        &\mathcal{V}(\mathcal{V}_2(\omega,C,\omega),\omega)-\mathcal{V}_2(\omega,C,\mathcal{V}(\omega,\omega))-\mathcal{V}_2(\omega,\mathcal{U}_2(C,\omega),\omega)-\mathcal{V}(\omega,\mathcal{V}_3(C,\omega,\omega))\\
        &+\mathcal{V}_3(\mathcal{U}_1(\omega,C),\omega,\omega)=0\,,\\
        &V(\mathcal{V}_3(C,\omega,\omega),\omega)-\mathcal{V}_3(C,\omega,\mathcal{V}(\omega,\omega))+\mathcal{V}_3(C,\mathcal{V}(\omega,\omega),\omega)+\mathcal{V}_3(\mathcal{U}_2(C,\omega),\omega,\omega)=0\,.
    \end{split}
\end{equation}
In Appendix \ref{app:complex} we rewrite the equations here-above in terms of symbols of operators. In Appendix \ref{app:vertices} we find a nontrivial solution. The idea is to look for regular vertices in the form of singular field-redefinitions. In other words, if the cocycle is formally trivial but the coboundary does not belong to the required functional class, the cocycle is nontrivial. The final result can be written as
\besubeqs
\begin{empheq}[box=\fbox]{align}
    \mathcal{V}_1(\omega,\omega,C)&: && +p_{12}\,S \int_{\Delta_2}\exp[\left(1-t_1\right) p_{01}+\left(1-t_2\right) p_{02}+t_1 p_{13}+t_2 p_{23}]\,, \\
    \mathcal{V}_2(\omega,C,\omega)&: &&\begin{aligned}
       -p_{13}\,S& \int_{\Delta_2}\exp[\left(1-t_2\right) p_{01}+\left(1-t_1\right) p_{03}+t_2 p_{12}-t_1 p_{23}]+\\
       -&p_{13}\,S \int_{\Delta_2}\exp[\left(1-t_1\right) p_{01}+\left(1-t_2\right) p_{03}+t_1 p_{12}-t_2 p_{23}]\,,
    \end{aligned}
    \\
    \mathcal{V}_3(C,\omega,\omega)&: && +p_{23}\,S \int_{\Delta_2}\exp[\left(1-t_2\right) p_{02}+\left(1-t_1\right) p_{03}-t_2 p_{12}-t_1 p_{13}]\,.
\end{empheq}
\esubeqs
Here, $\Delta_n$ is an $n$-dimensional simplex $t_0=0\leq t_1\leq ...\leq t_n\leq 1$. The nontriviality of the solution is proved in Appendix \ref{app:vertices}. There is an overall factor $S$
\begin{align}\label{juststar}
    S&= \exp[q_{01}+q_{02}+q_{03}+q_{12}+q_{13}+q_{23}]
\end{align}
that computes the star-product over all $y$ variables. In other words, the vertex factorizes 
\begin{align}
    \mathcal{V}_1(a(y)\otimes \bar{a}(\bry),b(y)\otimes \bar{b}(\bry),c(y)\otimes \bar{c}(\bry))&= a\star b\star c \otimes v_1(\bar{a},\bar{b},\bar{c})\,,
\end{align}
and similarly for the other vertices. 

\textit{Remark.} As it was pointed out in \cite{Gerasimenko:2021sxj,Sharapov:2018hnl,Sharapov:2022eiy}, constructing FDA's of higher spin gravities calls for an extension of the deformation quantization of Poisson manifolds to Poisson orbifolds, which is an open problem. Nevertheless, the traces of Kontsevich and of Shoikhet-Tsygan-Kontsevich formality are sometimes visible \cite{Sharapov:2017yde}. The key point in the proof of the formality theorems is to find the right configuration space and the right closed form on it, so that the proof amounts to a simple application of the Stokes theorem. As we show in Appendix \ref{app:vertices}, one can find a closed two-form $\Omega$, $d\Omega=0$, on $\Delta_3$ such that its integral over the four boundaries of the simplex reduces to the four terms in the equation for $\mathcal{V}_1$ and similarly for other vertices. In this regard let us note that the integral form is not unique. It arises as an integral over the configuration space of ordered points on a circle. With the help of translation invariance one can (gauge) fix the times of different points and also use the reflection symmetry of the circle. Altogether there are six different forms. 

\textit{Remark.} The cubic vertex has an interesting property: if we remove for a moment the matrix factors $\mathrm{Mat}_N$, make $y$ commutative (by taking the $\hbar=0$ limit after introducing $\hbar$ into the Moyal-Weyl star-product) and bring $\omega$'s and $C$ to the same $\omega\omega C$-ordering, we get zero:
\begin{align}
    \mathcal{V}_1(\omega,\omega,C)+\mathcal{V}_2(\omega,C,\omega)+\mathcal{V}_3(C,\omega,\omega)\Big|_{\hbar=0\,, N=1} &\equiv0\,.
\end{align}
This does not have to be the case. However, erasing matrix factors together with the commutative limit in $y$ must give a trivial vertex. Indeed, there is no such truncation of Chiral Theory. Therefore, the vertices we found enjoy some kind of minimality, giving zero whenever they should.

\textit{Remark.} It can be shown that $\mathcal{V}_{1,2,3}\neq0$. In all the cases considered before $C$ takes values in the (twisted)-adjoint representation of a higher spin algebra. This allows one to set $\mathcal{V}_{2,3}=0$ and choose $\mathcal{V}_{1}(a,b,c)=\phi(a,b)\star c$, where $\phi(a,b)$ is a certain Hochschild two-cocycle that deforms the higher spin algebra. Indeed, assuming $\mathcal{V}_{2,3}=0$ we find
\begin{align}\label{equivar}
    \mathcal{V}(\mathcal{V}_1(a,b,c),d)+\mathcal{V}_1(a,b,\mathcal{U}_2(c,d))&=0\,.
\end{align}
Here, $\mathcal{U}_2(c,d)=-c\star d$ and $\mathcal{V}(a,b)=a\star b$ in the previously studied cases. Therefore, setting $c=1$ leads to $\mathcal{V}_1(a,b,d)=\mathcal{V}_1(a,b,1)\star d$. Moreover, $\phi(a,b)=\mathcal{V}_1(a,b,1)$ turns out to be a Hochschild two-cocycle. For Chiral Theory this cannot be true. Indeed, it is easy to see that while in the second term of \eqref{equivar} $d$ has all of its indices contracted with $c$, the same indices are free in the first term, i.e. \eqref{equivar} cannot be satisfied. 
Therefore, we have to look for a solution with all $\mathcal{V}_{1,2,3}\neq0$, as we did.

\paragraph{Cubic Vertex $\boldsymbol{\mathcal{U}(\omega,C,C)}$. } The previously found vertex $\mathcal{V}(\omega,\omega,C)$ serves as a source for $\mathcal{U}(\omega,C,C)$. As before, we split it according to different orderings:
\begin{align}
    \mathcal{U}(\omega,C,C)=\mathcal{U}_1(\omega,C,C)+\mathcal{U}_2(C,\omega,C)+\mathcal{U}_3(C,C,\omega)   \,. 
\end{align}
There are six equations that can be obtained as various $\omega^2C^2$-terms after applying $d$ to \eqref{eq:chiraltheory}. We rewrite them in terms of symbols of operators in Appendix \ref{app:complex} and solve in Appendix \ref{app:vertices}. The final form of the solution reads:\footnote{A resemblance to some of the formulas in the literature \cite{Vasiliev:1988sa} is striking, of course. However, as different from \cite{Vasiliev:1988sa}, all vertices in the present paper are local and do not contain infinite (divergent) sums over different representations of the same interactions \cite{Boulanger:2015ova,Skvortsov:2015lja}. Therefore, we are constructing an actual theory rather than the most general ansatz for interactions compatible with symmetries. }
\besubeqs
\begin{empheq}[box=\fbox]{align}
    \mathcal{U}_1(\omega,C,C)&: && +p_{01}\,S \int_{\Delta_2}\exp[\left(1-t_2\right) p_{02}+t_2 p_{03}+\left(1-t_1\right) p_{12}+t_1 p_{13}]\,, \\
    \mathcal{U}_2(C,\omega,C)&: &&\begin{aligned}
       -p_{02}\,S& \int_{\Delta_2}\exp[t_2 p_{01}+\left(1-t_2\right) p_{03}-t_1 p_{12}+\left(1-t_1\right) p_{23}]+\\
       -&p_{02}\,S \int_{\Delta_2}\exp[t_1 p_{01}+\left(1-t_1\right) p_{03}-t_2 p_{12}+\left(1-t_2\right) p_{23}]\,,
    \end{aligned}
    \\
    \mathcal{U}_3(C,C,\omega)&: && +p_{03}\,S \int_{\Delta_2}\exp[\left(1-t_1\right) p_{01}+t_1 p_{02}+\left(t_2-1\right) p_{13}-t_2 p_{23}]\,,
\end{empheq}
\esubeqs
where $S$ is the star-product over $y$'s, \eqref{juststar}.

\subsection{Summary and Discussion}
\label{sec:}
The main result of this paper are the boxed formulas above that define vertices $\mathcal{V}(\omega,\omega)$, $\mathcal{U}(\omega,C)$, $\mathcal{V}(\omega,\omega,C)$, $\mathcal{U}(\omega,C,C)$. Altogether they satisfy the $L_\infty$-relations up to order $\mathcal{O}(C^2)$. These vertices determine both the free equations and the essential interactions of Chiral Theory. By essential we mean those that contribute to the cubic amplitude and which fully determine Chiral Theory. Let us recall that one can switch on very few higher-spin interactions and it is the consistency of the theory that will enforce the unique completion  \cite{Metsaev:1991mt,Metsaev:1991nb,Ponomarev:2016lrm}.  However, the covariantization may require more contact vertices, which is an interesting problem for the future. 

In Chiral Theory there is one dimensionful coupling constant, $l_P$, which is needed to compensate for higher powers of momenta in the vertices. The power of momenta equals the sum of the helicities, $\lambda_1+\lambda_2+\lambda_3$, of the fields that meet at the vertex. Given that the action of SDGR (with cosmological constant) contains  $d\omega^{A'B'}+\omega\fud{A'}{C'}\wedge \omega^{C'B'}$, it makes sense to assign mass dimension $1$ to all $\omega^{A'(2s-2)}$ and, hence, mass dimension zero to all $\Psi^{A'(2s)}$. Similarly, $e^{AA'}$ has dimension one. All $\omega^{A'(2s-2-k),A(k)}$ are expressed as derivatives of $\omega^{A'(2s-2)}$. It is then tempting to extend this to the whole $\omega$ and $C$. To recover $l_P$ we need to introduce it into $e^{AA'}$, e.g. $e^{AA'}_\mu \sim l_P^{-1} \sigma^{AA'}_\mu$ in Cartesian coordinates.

The dimensionless coupling $\kappa$ simply counts the orders of $\omega$ and $C$ in the perturbative expansion. In the light-cone gauge the expansion stops at the cubic terms. This does not have to be the case after covariantization. Let us compare the general structure of interactions in the light-cone gauge and in the FDA expanded over Minkowski vacuum $\omega_0=e$. We will be sketchy here. It is convenient to pack all positive helicity fields into $\Phi$ and all negative helicity fields plus scalar into $\Psi$. The action reads (very schematically)
\begin{align}\label{sketch}
    \mathcal{L}&= \Psi \square \Phi + c_{+++}\Phi\Phi\Phi+c_{++-}\Phi\Phi\Psi+c_{+--}\Phi\Psi\Psi\,,
\end{align}
where we drop the helicity labels and omit the detailed structure of interactions. The equations of motion would be
\begin{align}
    \square \Phi&= c_{++-}\Phi\Phi +c_{+--}\Phi\Psi\,, & \square\Psi&=c_{+++}\Phi\Phi+c_{++-}\Phi\Psi +c_{+--}\Psi\Psi\,.
\end{align}
This should be compared with ($D\equiv d-\omega_0$ is the background covariant derivative in the appropriate representations of the higher spin algebra)
\besubeqs\label{eq:chiraltheoryA}
\begin{align} 
    D\omega&= \mathcal{V}(\omega, \omega) +\mathcal{V}(\omega_0,\omega,C)+\mathcal{V}(\omega_0,\omega_0,C,C)\,,\\
    DC&= \mathcal{U}(\omega,C)+ \mathcal{U}(\omega_0,C,C) \,,
\end{align}
\esubeqs
where we indicated all terms that can potentially contribute to the cubic amplitude. We recall that $\omega$ carries positive helicity and, hence, is a cousin of $\Phi$, while $C$ contains both $\Psi$ and descendants of $\omega$. We show in appendix \ref{app:amplitude} that $\mathcal{V}(\omega, \omega)$ and $\mathcal{U}(\omega, C)$ give the correct amplitudes. There is a unique theory that has such amplitudes, which is a consistency check.

Another valuable consistency check is to restrict interactions to the spin-two and to the spin-one sectors to reproduce the recently obtained FDA's of SDYM and SDGR \cite{SDFDA}. To be precise, the restriction has to give FDA's that are quasi-isomorphic to those of SDYM and SDGR. Luckily, this exercise directly leads to the interactions of \cite{SDFDA}. The latter were found in the most minimal form, i.e. we have not introduced any nonlinear terms into the FDA beyond what is necessary, which fixes all field redefinitions. It is encouraging that the FDA of Chiral Theory is also minimal in this sense. 

By the same token the higher spin extensions of SDYM and SDGR \cite{Krasnov:2021nsq}, which were previously discovered as  contractions of Chiral Theory in \cite{Ponomarev:2017nrr}, must be consistent contractions of the present FDA as well. We note that in the latter two cases the FDA of this paper should provide a complete solution of the problem. Indeed, the actions of these two theories are schematically
\begin{align}
    \mathcal{L}&= \Psi \square \Phi +\Phi\Phi\Psi\,,
\end{align}
which is much simpler than the structure of interactions of Chiral Theory. Therefore, it is tempting to argue that we have determined all interaction vertices in these theories since this is the case for SDYM and SDGR. 

A very interesting observation made in \cite{Ponomarev:2017nrr} is that the coupling constants of Chiral Theory determine a certain (kinematic) algebra in the light-cone gauge and the product in this algebra is a remnant of the star-product. This statement covers all vertices. For the FDA at hand, it is the $\Phi\Phi\Psi$-vertex where the star-product structure is manifest. The other vertices correspond to the Chevalley-Eilenberg cocycles of the higher spin algebra. Nevertheless, according to \cite{Ponomarev:2017nrr}, what survives of these vertices in the light-cone gauge is the same star-product. It would be interesting to clarify this statement.

\section{Conclusions}
\label{sec:}
The main result of this paper is the covariant form that incorporates some essential interactions of Chiral Theory which was previously known in the light-cone gauge only. By essential we mean those interactions that, if present, unambiguously fix the theory. Technically, the result is the minimal model of Chiral Theory --- a Free Differential Algebra consistent to order $\mathcal{O}(C^2)$. 

The FDA of the present paper contains FDA's of SDYM, SDGR \cite{SDFDA} and of the higher spin extensions thereof \cite{Krasnov:2021nsq}. For these four cases the FDA should be complete. For Chiral Theory certain higher order vertices may still be required for formal consistency and covariantization. One can also look for supersymmetric extensions that would combine SDYM and SDGR and higher spin extensions thereof \cite{Devchand:1996gv} as well as for the full supersymmetric Chiral Theory \cite{Metsaev:2019dqt,Metsaev:2019aig}. 

Even though we found a covariant form for the essential interactions of Chiral Theory, there might still be an obstruction to getting the complete theory in a manifestly Lorentz invariant form if some of the interactions cannot be written with the help of the new field variables ($\omega^{A'(2s-2)}$ and $\Psi^{A'(2s)}$ as compared to the old $\Phi_{\mu_1...\mu_s}$) . In Appendix \ref{app:amplitude} we also show that the most problematic $V^{+--}$ amplitudes can also be reproduced. Independently of that, a simple extension \cite{Sharapov:2022new} of the cohomological arguments along the lines of \cite{Sharapov:2020quq} indicates that there are no obstructions to the FDA of this paper. Therefore, the complete Chiral Theory can be written in a manifestly Lorentz invariant form as an FDA.  

As is well-understood \cite{Barnich:2010sw,Grigoriev:2012xg,Grigoriev:2019ojp}, the minimal model of a (gauge) field theory contains all the essential information about the theory (local BRST cohomology), e.g. actions/counterterms, anomalies, conserved charges, deformations, etc. It is a very encouraging statement given that the differential $Q$ can be extracted from classical field equations rewritten as a Free Differential Algebra. Therefore, the results of this paper should help to address the problems where having a covariant form of the theory is an advantage, i.e. all of them. Chiral Theory was shown to be one-loop finite in the light-cone gauge \cite{Skvortsov:2018jea,Skvortsov:2020wtf,Skvortsov:2020gpn}, but extending these results to higher loop orders should be simpler within a covariant approach. It would also be interesting to look for exact solutions where generalizations of Ward/Penrose/ADHM \cite{Ward:1977ta,Penrose:1976js,Atiyah:1978ri} constructions to Chiral Theory together with its twistor formulation should be of great help.

\section*{Acknowledgments}
\label{sec:Aknowledgements}
We are grateful to Maxim Grigoriev, Yannick Herfray, Kirill Krasnov, Dmitry Ponomarev and Alexey Sharapov for useful discussions. This project has received funding from the European Research Council (ERC) under the European Union’s Horizon 2020 research and innovation programme (grant agreement No 101002551). The work was partially supported by the Fonds de la Recherche Scientifique --- FNRS under Grant No. F.4544.21.

\appendix

\section{Cubic Amplitude}
\label{app:amplitude}
A useful check for a given interaction is to compute the amplitude. The amplitudes of Chiral HiSGRA are known up to one-loop \cite{Skvortsov:2018jea,Skvortsov:2020wtf,Skvortsov:2020gpn}. We do not have to go that far and should just check if the cubic amplitude is nontrivial. Let us first construct the plane wave solutions. We recall that the free equations in Minkowski space read
\besubeqs\label{linearizeddataA}
\begin{align}
    d\omega &= e^{BB'}\bry_{B'} \pl_{B} \omega +H^{BB} \pl_{B}\pl_{B}C(y,\bry=0)\,, &
    d C&= e^{BB'}\pl_B \pl_{B'} C\,,
\end{align}
\esubeqs
where $\Psi(0,\bry)=C(0,\bry)$ describes negative helicity and $\omega(0,\bry)$ describes positive helicity:\footnote{Note that we use the Moyal-Weyl star-product without $i$. Therefore, the fields need to obey less natural reality conditions. This is not an obstacle to compute the amplitude. In particular, the plane wave exponents are taken without $i$ (for appropriate $x$). What matters is the helicity structure. }
\begin{align}
    \Psi^{A'(2s)}&= a_{-s}\, k^{A'}...k^{A'} \exp{[\pm x^{AA'} k_A k_{A'}]}\,,\\
    \omega^{A'(2s-2)}&= a_{+s}\, \frac{1}{(q^{C'}k_{C'})^{2s-1}} e^{BB'} k_B q_{B'} q^{A'} ...q^{A'}\exp{[\pm x^{AA'} k_A k_{A'}]}\,.
\end{align}
Here $a_\lambda$ is a normalization factor. Eq. \eqref{linearizeddataA} is solved by
\begin{align*}
    \omega(x|y,\bry)&=e^{BB'}\frac{k_B q_B'}{\bar{q}\bar{k}+\bar{y}\bar{q}}\exp(\pm x^{AA'}k_Ak_A'+yk)\,, & C(x|y,\bry)=\frac{1}{2}\exp(\pm x^{AA'}k_Ak_A'+yk+\bry\bar{k}) \,.
\end{align*}
Laplace transform allows us to rewrite the solution for $\omega(x|y,\bry)$ as
\begin{align*}
    \omega(x|,y,\bry)=e^{BB'}k_B q_B'\int_{0}^{\infty}d\omega\exp(\pm x^{AA'}k_Ak_A'+yk-(\bar{q}\bar{k}+\bar{y}\,\bar{q})\omega) \,.
\end{align*}
In order to compute  cubic amplitudes we can isolate the equation for $\omega^{A'(2s-2)}$ and $\Psi^{A'(2s)}$:
\begin{align} 
    D\omega&= \mathcal{V}(\omega, \omega)\Big|_{y=0}\,, &
    DC&= \mathcal{U}(\omega,C) \Big|_{y=0}\,.
\end{align}
Let us have a look at the first term $V(\omega, \omega)$ contracted with $\Psi^{A'(2s_1-2)}H_{A'A'}$ to get an on-shell cubic vertex:
\begin{align}
    \frac{1}{l!}\int \Psi_{A'(2s_1)}H^{A'A'}\, \omega^{B(l),A'(n)}\wedge \omega\fdu{B(l)}{,A'(m)}\,.
\end{align}
Here we assume $l+n=2s_2-2$, $m+l=2s_3-2$ and, of course, $m+n=2s_1-2$. The coefficient in front of the action originates from the star-product. Plugging in the on-shell plane-wave values for $\Psi$ and $\omega$ we find
\begin{align} \label{cubicVertex}
    V_{-s_1,+s_2,+s_3}&\sim \frac{1}{\Gamma[-s_1+s_2+s_3]} [12]^{-s_1+s_2-s_3}[23]^{s_2+s_3+s_1}[13]^{-s_1+s_3-s_2}\,,
\end{align}
which, up to normalization of each of the plane-waves, is the right structure for Chiral Theory. It corresponds to $\Psi\Phi\Phi$-vertex of sketch \eqref{sketch}. The presence of the simplest self-interaction $V_{-s,+s,+s}$ leads unambiguously to the Chiral Theory class since it requires all other spins (at least even) together with all other possible interactions that enter with weight $1/\Gamma[\lambda_1+\lambda_2+\lambda_3]$.

Similarly, we can extract the amplitudes corresponding to $\Phi\Phi\Phi$ and $\Phi\Phi\Psi$ vertices from $\mathcal{U}(\omega,C)$. Note that since $C$ contains both positive and negative (as well as zero) helicities, we get an access to two types of vertices. The final amplitude is
\begin{align*}
    V_{+s_1,\lambda_2,+s_3}&\sim 
       \frac{1}{\Gamma[s_1+\lambda_2+s_3]}[12]^{s_1+\lambda_2-s_3}[23]^{-s_1+\lambda_2+s_3}[13]^{s_1-\lambda_2+s_3}\,.
\end{align*}
Let us also comment on the possibility to reproduce $V_{+s_1,-s_2,-s_3}$ amplitudes, $s_1-s_2-s_3>0$. From the standard covariant approach vantage point, where the dynamical variables are $\Phi_{\mu_1...\mu_s}$, these vertices are the most problematic ones \cite{Conde:2016izb}. They cannot be written at all as local expressions. Fortunately, it is easy to write down the candidate on-shell cubic vertices in terms of the new variables, where the dynamical fields are $\omega^{A'(2s-2)}$ and $\Psi^{A'(2s)}$. For example, any of the following two expressions 
\begin{align*}
   &  \omega^{A'(s_1-2)}\Psi_{A(k)B,A'(m),B'}\Psi\fud{A(k)}{A'(n)}\hat{h}^{BB'}\,,  && \omega^{A'(s_1-2)}\Psi_{A(k)B,A'(m)}\Psi\fud{A(k)}{A'(n)B'}\hat{h}^{BB'},
\end{align*}
leads to the correct amplitude
\begin{align*}
    \begin{aligned}
       [12]^{s_1-s_2+s_3}[13]^{s_1+s_2-s_3}[23]^{-s_1-s_2-s_3}
    \end{aligned}
\end{align*}
Therefore, all possible types of cubic vertices/amplitudes present in Chiral Theory can be written in a manifestly Lorentz invariant way. This eliminates the very last obstruction and we can claim that Chiral Theory admits a manifestly Lorentz invariant formulation.

\section{Coadjoint vs. twisted-adjoint}
\label{app:coadjoint}
Let us make a historical remark on representations of higher spin symmetries. It was known since \cite{Vasiliev:1986td} that the FDA of free massless fields in (anti)-de Sitter space contains the following subsystem
\begin{align}
    \nabla C&= e^{AA'}(y_A \bry_{A'} -\pl_A \pl_{A'}) C(y,\bry)\,.
\end{align}
It splits according to spin into an infinite set of (still infinite) subsystems. For a given $s>0$ the subsystem splits further into one for helicity $+s$ and another one for helicity $-s$. The very first equations in these subsystems are equivalent to \cite{Penrose:1965am}
\begin{align}
    \nabla^\fdu{B}{A'} C^{BA(2s-1)}&=0\,, &\nabla\fud{A}{B'} C^{B'A'(2s-1)}&=0\,.
\end{align}
Operator $P_{AA'}=(y_A \bry_{A'} -\pl_A \pl_{A'})$ realizes the action of $(A)dS_4$ translations, which commute to a Lorentz transformation. Since the equations are assumed to be derived by linearizing a nonlinear theory, where the higher spin symmetry is manifest, it is important to understand where such $P_{AA'}$ can come form. It originates from the twisted-adjoint action \cite{Vasiliev:1999ba}:
\begin{align}\label{twad}
    a(f)&= a\star f-f\star \tilde{a}\,,
\end{align}
where $\tilde{a}$ is an automorphism of the Weyl algebra that flips the sign of $\bry$, $\tilde{a}(\bry)=a(-\bry)$. In fact, the action arises as a typical coadjoint action. Indeed, there is a nondegenerate pairing between $A_1$ and $A_1^\star$: $\langle a|f\rangle=\tr[a\star f]=\tr [f\star \tilde{a}]$, where $\tr[a]=a(\bry=0)$. The canonical bimodule structure of the higher spin algebra on itself (left/right actions) induces the twisted-adjoint representation \eqref{twad} on the dual module. What the results of the present paper show is that the coadjoint interpretation seems to be correct even for such a strange case as Chiral Theory, while the twisted-adjoint interpretation is no longer valid. 

\section{Operator calculus}
\label{app:}
As was already sketched at the beginning of Section \ref{sec:FDA}, we work with poly-differential operators that are represented as symbols. Let us illustrate all operations with $\bry$ and $\pl_{A'}^{\bry_i}\equiv p^i_{A'}$. The translation operator is $\exp{[\bry\cdot p_i]}f(\bry_i)= f(\bry_i+\bry)$. Operators acting on $n$ functions $a_i(\bry)$ are understood as functions of $p_0=\bry$, $p_1=\bar{\pl}_1$, ..., $p_n=\bar{\pl}_n$:
\begin{align}
V(a_1,...,a_n)=v(\bry,\bar{\pl}_1,...,\bar{\pl}_2)a_1(\bry_1)...a_n(\bry_n)\Big|_{\bry_i=0}
\end{align}
Therefore, the commutative product $f(\bry)g(\bry)$ and the Moyal-Weyl star-product $f(y)\star g(y)$ are represented by the following symbols:\besubeqs
\begin{align}
    \exp[p_0 \cdot p_1+p_0 \cdot p_2]\equiv \exp[p_{01}+p_{02}]\,, \\
    \exp[q_0 \cdot q_1+q_0 \cdot q_2+q_1 \cdot q_2]\equiv \exp[q_{01}+q_{02}+q_{12}]\,.
\end{align}
\esubeqs
Then we need the following identifications for symbols of the operators:
\begin{align*}
a_1\star V(a_2,...,a_{n+1})&\rightarrow  v(q_0+q_1,q_2,...,q_{n+1})e^{+ q_{0}\cdot q_1}\,,\\
V(a_1,...,a_n)\star a_{n+1}&\rightarrow v(q_0-q_{n+1},q_1,...,q_n)e^{+ q_{0}\cdot q_{n+1}}\,,\\
V(a_1,...,a_k\star a_{k+1},...,a_{n+1})&\rightarrow v(q_0,...,q_{k-1},q_k+q_{k+1},q_{k+2},...,q_{n+1})e^{+ q_{k}\cdot q_{k+1}}\,,
\end{align*}
\begin{align*}
a_1 V(a_2,...,a_{n+1})&\rightarrow  v(p_0,p_2,...,p_{n+1})e^{+ p_{0}\cdot p_1}\,,\\
V(a_1,...,a_n)a_{n+1}&\rightarrow v(p_0,p_1,...,p_n)e^{+ p_{0}\cdot p_{n+1}}\,,\\
V(a_1,...,a_k a_{k+1},...,a_{n+1})&\rightarrow v(p_0,...,p_{k-1},p_k+p_{k+1},p_{k+2},...,p_{n+1})\,,
\end{align*}
\begin{align*}
u_1(a_1, V(a_2,...,a_{n+1}))&\rightarrow  v(p_0+p_1,p_2,...,p_{n+1})\,,\\
u_1(V(a_1,...,a_n),a_{n+1})&\rightarrow v(-p_{n+1},p_1,...,p_n)e^{+ p_{0}\cdot p_{n+1}}\,,\\
V(a_1,...,u_1(a_k, a_{k+1}),...,a_{n+1})&\rightarrow v(p_0,...,p_{k-1},p_{k+1},p_{k+2},...,p_{n+1})e^{+p_{k}\cdot p_{k+1}}\,,
\end{align*}
\begin{align*}
u_2(a_1, V(a_2,...,a_{n+1}))&\rightarrow  v(-p_1,p_2,...,p_{n+1})e^{+ p_{0}\cdot p_1}\,,\\
u_2(V(a_1,...,a_n),a_{n+1})&\rightarrow v(p_0+p_{n+1},p_1,...,p_n)\,,\\
V(a_1,...,u_2(a_k, a_{k+1}),...,a_{n+1})&\rightarrow v(p_0,...,p_{k-1},p_k,p_{k+2},...,p_{n+1})e^{-p_{k}\cdot p_{k+1}}\,,
\end{align*}
where we defined
\begin{align}
        &u_1(a,b)=\exp{[p_{02}+p_{12}]}\,, &
        &u_2(a,b)=\exp{[p_{01}-p_{12}]}\,.
\end{align}

\section{Cochain complex}
\label{app:complex}
For completeness let us rewrite the $L_\infty$-relations in terms of symbols of operators. We do so for the $\bry$-part only since the dependence on $y$ is captured by the star-product and factorizes out. The l.h.s. of the equations for $\mathcal{V}_{1,2,3}$ read:
{\footnotesize
\begin{align*}
    &-e^{p_{01}} V_1\left(p_0,p_2,p_3,p_4\right)-V_1\left(p_0,p_1,p_2+p_3,p_4\right)+V_1\left(p_0,p_1+p_2,p_3,p_4\right)+e^{p_{34}} V_1\left(p_0,p_1,p_2,p_4\right) \,, \\
    &-e^{p_{01}} V_2\left(p_0,p_2,p_3,p_4\right)+e^{p_{04}} V_1\left(p_0,p_1,p_2,p_3\right)-e^{-p_{34}} V_1\left(p_0,p_1,p_2,p_3\right)+V_2\left(p_0,p_1+p_2,p_3,p_4\right)-e^{p_{23}} V_2\left(p_0,p_1,p_3,p_4\right) \,, \\
   & -e^{p_{01}} V_3\left(p_0,p_2,p_3,p_4\right)+e^{p_{04}} V_2\left(p_0,p_1,p_2,p_3\right)-V_2\left(p_0,p_1,p_2,p_3+p_4\right)+e^{-p_2\cdot p_3} V_2\left(p_0,p_1,p_2,p_4\right)+e^{p_{12}} V_3\left(p_0,p_2,p_3,p_4\right) \,, \\
   & e^{p_{04}} V_3\left(p_0,p_1,p_2,p_3\right)-e^{-p_{12}} V_3\left(p_0,p_1,p_3,p_4\right)-V_3\left(p_0,p_1,p_2,p_3+p_4\right)+V_3\left(p_0,p_1,p_2+p_3,p_4\right) \,.
\end{align*}}\noindent
Similarly, for $\mathcal{U}_{1,2,3}$ we find 
{\footnotesize
\begin{align*}
  & U_1\left(p_0,p_1+p_2,p_3,p_4\right)-U_1\left(p_0+p_1,p_2,p_3,p_4\right)-e^{p_{23}} U_1\left(p_0,p_1,p_3,p_4\right)+e^{p_{04}} V_1\left(-p_4,p_1,p_2,p_3\right) \,, \\
    &e^{-p_{23}} U_1\left(p_0,p_1,p_2,p_4\right)-U_2\left(p_0+p_1,p_2,p_3),
    p_4\right)-e^{p_{34}} U_1\left(p_0,p_1,p_2,p_4\right)+e^{p_{12}} U_2\left(p_0,p_2,p_3,p_4\right)+e^{p_{04}} V_2\left(-p_4,p_1,p_2,p_3\right),\\
   & e^{-p_{34}} U_1\left(p_0,p_1,p_2,p_3\right)-U_1\left(p_0+p_4,p_1,p_2,p_3\right)-U_3\left(p_0+p_1,p_2,p_3,p_4\right)+e^{p_{12}} U_3\left(p_0,p_2,p_3,p_4\right),\\
    &-e^{-p_{12}} U_3\left(p_0,p_1,p_3,p_4\right)+e^{-p_{34}} U_2\left(p_0,p_1,p_2,p_3\right)-U_2\left(p_0+p_4,p_1,p_2,p_3\right)+e^{p_{23}} U_3\left(p_0,p_1,p_3,p_4\right)-e^{p_{01}} V_2\left(-p_1,p_2,p_3,p_4\right),\\
   & -e^{-p_{12}} U_2\left(p_0,p_1,p_3,p_4\right)+U_2\left(p_0,p_1,p_2+p_3,p_4\right)-e^{p_{34}} U_2\left(p_0,p_1,p_2,p_4\right)-e^{p_{01}} V_1\left(-p_1,p_2,p_3,p_4\right)+e^{p_{04}} V_3\left(-p_4,p_1,p_2,p_3\right),\\
   & -e^{-p_{23}} U_3\left(p_0,p_1,p_2,p_4\right)+U_3\left(p_0,p_1,p_2,p_3+p_4\right)-U_3\left(p_0+p_4,p_1,p_2,p_3\right)-e^{p_{01}} V_3\left(-p_1,p_2,p_3,p_4\right) \,.
\end{align*}
}\noindent
When looking for nontrivial solutions, it is important to understand which ones are trivial. The latter are given by field redefinitions that act as follows on  $\mathcal{V}_{1,2,3}$
\begin{align*}
    \delta V_1&=e^{p_{01}} g_1\left(p_0,p_2,p_3\right)-g_1\left(p_0,p_1+p_2,p_3\right)+e^{p_{23}} g_1\left(p_0,p_1,p_3\right)\,,\\
   \delta V_2&= e^{p_{01}} g_2\left(p_0,p_2,p_3\right)+e^{p_{03}} g_1\left(p_0,p_1,p_2\right)-e^{-p_{23}} g_1\left(p_0,p_1,p_2\right)-e^{p_{12}} g_2\left(p_0,p_2,p_3\right)\,,\\
   \delta V_3&= e^{p_{03}} g_2\left(p_0,p_1,p_2\right)+e^{-p_{12}} g_2\left(p_0,p_1,p_3\right)-g_2\left(p_0,p_1,p_2+p_3\right)\,,
\end{align*}
and on $\mathcal{U}_{1,2,3}$
\begin{align*}
    \delta U_1&=h\left(p_0+p_1,p_2,p_3\right)-e^{p_{12}} h\left(p_0,p_2,p_3\right)+e^{p_{03}} g_1\left(-p_3,p_1,p_2\right)\,,\\
   \delta U_2&= e^{-p_{12}} h\left(p_0,p_1,p_3\right)-e^{p_{23}} h\left(p_0,p_1,p_3\right)-e^{p_{01}} g_1\left(-p_1,p_2,p_3\right)+e^{p_{03}} g_2\left(-p_3,p_1,p_2\right)\,,\\
   \delta U_3&= e^{-p_{23}} h\left(p_0,p_1,p_2\right)-h\left(p_0+p_3,p_1,p_2\right)-e^{p_{01}} g_2\left(-p_1,p_2,p_3\right)\,,
\end{align*}
It can easily be checked that the redefinitions lead to solutions of the equations. The expressions above define a particular realization of the Chevalley-Eilenberg complex, but we do not extend the action of the differential to cochains with more arguments. At the bottom level we find
\begin{align*}
    \delta g_1&= e^{p_{12}} \xi \left(p_0,p_2\right)-e^{p_{01}} \xi \left(p_0,p_2\right)\,,\\
    \delta g_2&= e^{p_{02}} \xi \left(p_0,p_1\right)-e^{-p_{12}} \xi \left(p_0,p_1\right)\,,\\
    \delta h&= e^{p_{02}} \xi \left(-p_2,p_1\right)-e^{p_{01}} \xi \left(-p_1,p_2\right)\,,
\end{align*}
which leads to redefinitions that yield vanishing vertices. 

\section{Vertices}
\label{app:vertices}
In order to find nontrivial cubic vertices we employ a number of ideas, see also \cite{Vasiliev:1988sa,Sharapov:2017yde} that were used for inspiration. Firstly, Lorentz symmetry has to be preserved, i.e., in practice, we cannot mix primed and unprimed indices. The higher spin algebra is the tensor product of two algebras, which via the K{\"u}nneth theorem suggests to look for the two-cocycle as a tensor product of two, one of them being trivial. The free equations, in particular the boundary condition for $\mathcal{V}(e,e,C)$, reveal that something interesting should happen on the $\bry$ side. Therefore, for homogeneous arguments $a(y,\bry)=a(y)\otimes \bar{a}(\bry)$, etc. we assume that all vertices have the star-product over the $y$-dependent factors:
\begin{align}
    \mathcal{V}_1(a(y)\otimes \bar{a}(\bry),b(y)\otimes \bar{b}(\bry),c(y)\otimes \bar{c}(\bry))&= a\star b\star c \otimes v_1(\bar{a},\bar{b},\bar{c})\,.
\end{align}
As a result, all terms in the cocycle equations have the same overall factor for the $y$-dependence and we can concentrate on $\bry$ only. The cocycle conditions for the $\bry$-part are collected in Appendix \ref{app:complex}. 

Now, we need to solve the equations in Appendix \ref{app:complex}. It is clear that the solution should contain some $\exp[p_{ij}]$-factors, otherwise they cannot cancel the $\exp{[p_{ij}]}$ already present in the cocycle condition. The boundary condition for $\mathcal{V}(\omega,\omega,C)$ restrict the exponents a little bit. For example, we cannot allow for $\exp{p_{03}}$ in $\mathcal{V}_1(\omega,\omega,C)$. The crucial step is to look for $\mathcal{V}$ and $\mathcal{U}$ as singular field redefinitions, i.e. we look for $g_{1,2}$ and $h$, see Appendix \ref{app:complex}. For any $g_{1,2}$ and $h$, the vertices solve the cocycle equations. We just need to make sure that (i) the vertices are regular, i.e. Taylor expandable in $p_{ij}$; (ii) the redefinitions themselves, i.e. $g_{1,2}$ and $h$, are irregular. Irregular field redefinitions are not allowed. Therefore, if (i) and (ii) are satisfied, we have a nontrivial cocycle. Let us note that the singularity of $g_{1,2}$ and $h$ must be essential and cannot be removed with the help of "redefinitions for redefinitions" with $\xi$. Looking for singular redefinitions is more economic than looking for nontrivial vertices since they depend on less arguments. Long story short, we arrived at the following redefinitions:
\besubeqs
\begin{align}
    g_1&= \frac{p_{01} e^{p_{12}}}{p_{02} \left(p_{01}-p_{12}\right)}-\frac{e^{p_{01}} p_{01}}{p_{02} \left(p_{01}-p_{12}\right)}\,,\\
    g_2&= \frac{e^{p_{02}} p_{02}}{p_{01} \left(p_{02}+p_{12}\right)}-\frac{p_{02} e^{-p_{12}}}{p_{01} \left(p_{02}+p_{12}\right)}\,,\\
    h&= \frac{e^{p_{01}} p_{01}}{p_{12} \left(p_{01}-p_{02}\right)}-\frac{e^{p_{02}} p_{02}}{p_{12} \left(p_{01}-p_{02}\right)}\,.
\end{align}
\esubeqs
The vertices, which we do not write here as fractions, have a similar structure and their regularity is not obvious. It is very important to take advantage of the Fierz/Schouten/Pl{\"u}cker identities 
\begin{equation}
    (a\cdot b)(c\cdot d)+(b\cdot c)(a\cdot d)-(a\cdot c)(b\cdot d)=0\,,
\end{equation}
which are a consequence of the fact that any three vectors in two dimensions are linearly dependent. Still, the regularity is not manifest. One can prove it by showing that the numerator and denominator have the same zeros. 

A more convenient way to make the regularity manifest to write the vertices as integrals over the $2d$-simplex, as in the main text. The nontriviality of the cocycles is then less obvious. A simple way to check if the cocycle is nontrivial is to extract the boundary condition $\mathcal{V}(e,e,C)$ since this part cannot be redefined away. Therefore, once the boundary condition is satisfied we can be certain that the cocycle is worthy. It would be interesting to compute the Chevalley-Eilenberg cohomology following the techniques of \cite{Sharapov:2020quq}, which would give a rigorous answer regarding the number of independent vertices within the covariant approach. 

Given the relation between the algebraic structures of higher spin gravities and deformation quantization and formality, it is also possible to recast the proof into the familiar language of Stokes theorem. For example, to check that the equation for $\mathcal{V}_1$ is satisfied we can construct a closed two-form $\Omega_1$
\begin{align}
\begin{aligned}
        \Omega_1&= ( p_{12}\, dt_1 \wedge dt_2 +p_{23}\, dt_2\wedge dt_3 + p_{13}\,dt_1 \wedge dt_3)F_1 \,,\\
    F_1&=\exp{[\left(1-t_1\right) p_{01}+\left(1-t_2\right) p_{02}+\left(1-t_3\right) p_{03}+t_1 p_{14}+t_2 p_{24}+t_3 p_{34}]}\,.
\end{aligned}
\end{align}
With the help of Stokes theorem we get
\begin{align}
    0&=\int_{\Delta_3} d\Omega_1= \int_{\pl\Delta_3} \Omega_1\,.
\end{align}
There are four boundaries that correspond to "collisions of points" on the circle: $t_1=0$, $t_1=t_2$, $t_2=t_3$ and $t_3=1$. It can be seen that $\Omega_1$ at these boundaries reduces to exactly the four terms in the equation for $\mathcal{V}_1$. Similar arguments are true for the rest of the equations. The closed two-form for the other equations are
\besubeqs
\begin{align}
    \begin{aligned}
        \Omega_2&= ( p_{12}\, dt_1 \wedge dt_2 +p_{24}\, dt_2\wedge dt_3 + p_{14}\,dt_1 \wedge dt_3)F_2 \,,\\
        &- ( p_{14}\, dt_1 \wedge dt_2 -p_{12}\, dt_2\wedge dt_3 + p_{24}\,dt_1 \wedge dt_3)F_3 \,, \\
    F_2&=\exp{[\left(1-t_1\right) p_{01}+\left(1-t_2\right) p_{02}+\left(1-t_3\right) p_{04}+t_1 p_{13}+t_2 p_{23}-t_3 p_{34}]} \,, \\
    F_3&=\exp{[\left(1-t_2\right) p_{01}+\left(1-t_3\right) p_{02}+\left(1-t_1\right) p_{04}+t_2 p_{13}+t_3 p_{23}-t_1 p_{34}]} \,,
    \end{aligned}
\end{align}
for the second and for the third we need
\begin{align}
        \begin{aligned}
        \Omega_3&= - ( p_{34}\, dt_1 \wedge dt_2 +p_{13}\, dt_2\wedge dt_3 + p_{14}\,dt_1 \wedge dt_3)F_4\,, \\
        &+ ( p_{14}\, dt_1 \wedge dt_2 -p_{34}\, dt_2\wedge dt_3 + p_{13}\,dt_1 \wedge dt_3)F_5 \,, \\
    F_4&=\exp{[\left(1-t_3\right) p_{01}+\left(1-t_2\right) p_{03}+\left(1-t_1\right) p_{04}+t_3 p_{12}-t_2 p_{23}-t_1 p_{24}]} \,, \\
    F_5&=\exp{[\left(1-t_1\right) p_{01}+\left(1-t_3\right) p_{03}+\left(1-t_2\right) p_{04}+t_1 p_{12}-t_3 p_{23}-t_2 p_{24}]} \,.
    \end{aligned}
\end{align}
Note that the 2nd and 3rd equations have more terms since they mix vertices with different orderings and for this reason two exact forms are required. There is some mutual cancellation between them. For the last equation we have 
\begin{align}
\begin{aligned}
        \Omega_4&= - ( p_{34}\, dt_1 \wedge dt_2 +p_{23}\, dt_2\wedge dt_3 + p_{24}\,dt_1 \wedge dt_3)F_6\,, \\
    F_6&=\exp{[\left(1-t_3\right) p_{02}+\left(1-t_2\right) p_{03}+\left(1-t_1\right) p_{04}-t_3 p_{12}-t_2 p_{13}-t_1 p_{14}]} \,.
\end{aligned}
\end{align}
\esubeqs
The two-forms can be understood as follows: the first term in $\Omega_2$ is $\Omega_1$ with the labels $3$ and $4$ swapped, whereas the second term arises from cyclic permutation of the labels ($1234\rightarrow 2341$). Subsequently, $\Omega_4$ is the mirror image of $\Omega_1$, i.e. $1234\rightarrow 4321$, and $\Omega_3$ is the mirror image of $\Omega_2$. This can be understood from the different orderings of $\omega,\omega,\omega,C$ in the equations.

\footnotesize
\providecommand{\href}[2]{#2}\begingroup\raggedright\endgroup

\end{document}